\documentclass[11pt,a4paper]{article}
\usepackage[dvips]{graphicx}
\usepackage{amsmath}
\usepackage{cite}
\usepackage{amsfonts}
\usepackage{amssymb}
\begin{document}

\title{Network synchronization:\\
Spectral versus statistical properties}
\author{Fatihcan M. Atay%
\thanks{
Max Planck Institute for Mathematics in the Sciences,
 Inselstr.~22, D-04103 Leipzig, Germany (\texttt{atay@member.ams.org})
 } 
\and
T{\"{u}}rker B\i y\i ko\u{g}lu% 
\thanks{I\c{s}{\i}k University, Faculty of Arts and Sciences, Department of Mathematics,
 \c{S}ile 34980, Istanbul, Turkey
 (\texttt{biyikoglu@inma.ucl.ac.be})}
\and
 J{\"{u}}rgen Jost%
\thanks{Max Planck Institute for Mathematics in the Sciences,
 Inselstr.~22, D-04103 Leipzig, Germany 
 (\texttt{jost@mis.mpg.de})}
}

\date{}

\maketitle

\begin{abstract}
We consider synchronization of weighted networks, possibly with asymmetrical connections.
We show that the synchronizability of the networks cannot be directly inferred from
their statistical properties. Small local changes in the network structure can
sensitively affect the eigenvalues relevant for synchronization, while the
gross statistical network properties remain essentially unchanged.
Consequently, commonly used statistical properties, including the degree
distribution, degree homogeneity, average degree, average distance, degree correlation, and
clustering coefficient, can fail to characterize the synchronizability of networks.
\end{abstract}

\bigskip 

\bigskip

\textsc{Preprint.} Final version in \textit{Physica D,} 224:35-41, 2006.

\section{Introduction}

The description and classification of complex networks are often based on
their statistical properties, such as the degree distribution, average degree,
average distance, clustering coefficient, and degree correlations, among
others \cite{Albert-Barabasi-RevModPhys02,Dorogovtsev03,Newman03}. Indeed, the
starting point for the recent explosion of interest in complex networks can be
traced to the observation that real networks have degree distributions that
are much different from those of classical random graphs
\cite{Barabasi-Albert99}. On the other hand, the dynamics of processes defined
on networks are intimately related to the spectrum of an appropriate
connection operator. A prototypical example is chaos synchronization
\cite{Pecora90}, which crucially depends on the extremal eigenvalues of the
graph Laplacian \cite{Pecora98,Jost02,Li-Chen03}. This raises the natural
question of if and how the statistical properties of a network are related to
its spectral properties. Many recent papers have investigated various facets
of this relation. For example, some papers have reported correlations between
network synchronizability and degree homogeneity
\cite{Nishikawa03,Motter05,Motter-PRE05}, clustering coefficient
\cite{Wu-Wang06}, degree correlations \cite{diBernardo-arxiv05}, average
degree, degree distribution, and so on \cite{Hong04}. In some cases the
observed correlations can point in opposite directions; for instance,
\cite{Nishikawa03} finds that increasing the degree homogeneity improves
synchronizability, whereas \cite{Hong04} and \cite{diBernardo-arxiv05} report
cases of better synchronizability for decreased homogeneity. Similarly, adding
a few shortcut links to a sparse lattice is known to decrease the
characteristic path length and improve synchronizability at the same time
\cite{Barahona02,Hong02}, although another study showed that better
synchronization can result despite increased average distance
\cite{Nishikawa03}. Clearly, in view of the multitude of graph
characteristics, it can be difficult to translate the numerically observed
correlations into causal relations. Rigorous mathematical methods are
important for investigating the relations between different network
properties. The present paper provides a step in this direction. We give a
mathematical argument which shows that many statistical network properties do
not suffice to determine synchronizability. We present examples showing that
networks with the same statistical properties can have very different
synchronization characteristics. The results establish that the spectral
properties of networks are not simply derivable from statistical properties,
and should therefore hold their own place within the list of intrinsic network features.

\section{Spectral properties and structure}

\label{sec:structure}

Consider a network of $n$ nodes (vertices), with links (edges) between certain
pairs of nodes, which may additionally carry weights indicating the strength
of the relation they represent. We use the nonnegative numbers $a_{ij}$ to
denote the weight on the link from the $j$th node to the $i$th node, where
$a_{ij}=0\ $if and only if there is no link from $j$ to $i$. In general
$a_{ij}\neq a_{ji}$, although symmetric connections arise naturally in many
common models.
The degree of the vertex $i$ is defined as $\deg(i)=\sum_{j\in V}a_{ij}$.
Unweighted networks appear as a special case where each link carries a weight
of 1---in this case $A=[a_{ij}]$ is the usual adjacency matrix, and the degree
of a vertex is the number of its neighbors.

The Laplacian matrix is defined by $L=D-A$, where $D$ denotes the diagonal matrix of vertex
degrees. 
In case $A$ is symmetric, the Laplacian
is a symmetric and positive semidefinite matrix. Therefore, it has real and
nonnegative eigenvalues, which we order as $\lambda_{1}\leq\lambda_{2}%
\leq\cdots\leq\lambda_{n}$ (counting multiplicities), and an orthogonal set of
eigenvectors $\{\mathbf{u}_{1},\dots,\mathbf{u}_{n}\}$ which form a basis for
$\mathbb{R}^{n}$. Since the row sums of $L$ are zero, the smallest eigenvalue
$\lambda_{1}$ is always zero, and the corresponding eigenvector is
$\mathbf{1}=(1,1,\dots,1)$. The multiplicity of the zero eigenvalue 
equals the number of connected components of the network. In particular, the
second eigenvalue $\lambda_{2}$ is nonzero if and only if the network is
connected, which is one of the most fundamental relations between the network
structure and the spectrum of the connection operator.

For undirected networks, simple bounds can be
given for the eigenvalues in terms of the vertex degrees, which provide
further insight into the relation between the structural and spectral
properties. Let $d_{\min}$ and $d_{\max}$ denote, respectively, the smallest
and the largest degree, and let $\lambda_{\max}$
be the largest eigenvalue of the Laplacian. Then the following estimates
are well-known $($e.g.~\cite{Mohar04}):%
\begin{equation}
\lambda_{2}\leq\frac{n}{n-1}d_{\min}\leq\frac{n}{n-1}d_{\max}\leq\lambda
_{\max}\leq2d_{\max}.\label{est-degree}%
\end{equation}
Similarly, in terms of the average degree $d_{\mathrm{avg}}$ it can be shown
that
\begin{equation}
d_{\mathrm{avg}}<\lambda_{\max};\label{est-degree2}%
\end{equation}
see e.g.~\cite{PRE05}. Note that the second eigenvalue does not have a simple
bound from below in terms of the vertex degrees. This observation will be
important later on, as we show that $\lambda_{2}$ can indeed be arbitrarily
small among a class of networks having the same vertex degrees.

\section{Spectral properties and synchronization}

\label{sec:sync} The nodes of a network are often dynamical systems evolving
according to certain rules, and the links represent their pairwise
interaction. A typical interaction type is diffusion, which forms the
prototypical example where synchronization is observed \cite{Pikovsky-book01},
and naturally gives rise to the Laplacian
operator $L$. It is thus no coincidence that the dynamical properties are
closely related to the structural properties of the network. To focus on a
well-known example, we consider the case of the so-called coupled map lattice
\cite{Kaneko-book93}%
\begin{equation}
x_{i}(t+1)=f(x_{i}(t))+\sum_{j=1}^{n}a_{ij}\left[  f(x_{j}(t))-f(x_{i}%
(t))\right]  \label{network}%
\end{equation}
which we have written in a slightly more general form by allowing individual
weights $a_{ij}\geq0$ along the links instead of a common coupling strength
for the whole network. Denoting $\mathbf{x}=(x_{1},\dots,x_{n})$ and
$F(\mathbf{x})=(f(x_{1}),\dots,f(x_{n}))$, the system (\ref{network})
can be written in vector form%
\begin{equation}
\mathbf{x}(t+1)=(I-L)F(\mathbf{x}(t)).\label{vec}%
\end{equation}

The network (\ref{network}) is said to synchronize if $\lim_{t\rightarrow
\infty}|x_{i}(t)-x_{j}(t)|=0$ for all $i,j$ whenever the initial conditions
belong to some appropriate open set\footnote{For chaotic synchronization there
are some subtleties regarding the nature of the attraction and the open set of
initial conditions; the interested reader is referred to
\cite{time-varying-sync} for a clarification of such issues. The details,
however, will not be important for the derivation presented here.}. In this
case, the system asymptotically approaches a synchronous state, where each
node exhibits the same time evolution, $x_{i}(t)=s(t)$ for all $i$, or
$\mathbf{x}(t)=\mathbf{1}s(t)$. It follows from (\ref{network})
that
$
s(t+1)=f(s(t));\label{s}%
$
i.e., the behavior of the nodes in the synchronous state is identical to
their behavior in isolation\footnote{Here we neglect any coupling delays in
the network. The synchronous solutions can be markedly different when delays
are introduced; see \cite{Atay-PRL04,Atay-Complexity04}.}. In this paper we
focus on chaotic synchronization, that is, the case when $f$ has a compact
chaotic attractor $\mathcal{A}$ and $s$ represents some dense (and necessarily
unstable) orbit in $\mathcal{A}$. Assuming that $f$ is continuously
differentiable, small perturbations $u$ about the solution $s(t)$ are
governed by the equation $u(t+1)=f^{\prime}(s(t))u(t)$, which has the
solution%
\[
u(t)=u(0)\prod_{k=0}^{t-1}f^{\prime}(s(k)).
\]
Hence, the condition for local asymptotic stability of $s(t)$ is that
\begin{equation}
\lim_{t\rightarrow\infty}\prod_{k=0}^{t-1}|f^{\prime}(s(k))|=0.\label{stab}%
\end{equation}
While (\ref{stab}) would not hold for any solution $s$ inside a
chaotic attractor, it is always possible to find some sufficiently
large number $\alpha$ such that
\begin{equation}
\lim_{t\rightarrow\infty}\prod_{k=0}^{t-1}e^{-\alpha}|f^{\prime}%
(s(k))|=0.\label{stab-a}%
\end{equation}
In fact, it is easy to see that (\ref{stab-a}) holds for all $\alpha$
satisfying
\begin{equation}
\alpha>\mu\triangleq\lim_{t\rightarrow\infty}\frac{1}{t}\sum_{k=0}^{t-1}%
\log|f^{\prime}(s(k))|;\label{lyap}%
\end{equation}
where $\mu$ denotes the Lyapunov exponent.

Synchronization of coupled map lattices has been studied in more
or less general forms; e.g., \cite{Pikovsky-book01,Jost02,Lu-Chen04}. To find
the corresponding conditions, one considers small perturbations $\mathbf{u}%
(t)=\mathbf{x}(t)-\mathbf{1}s(t)$, which are governed by the 
variational equation
\[
\mathbf{u}(t+1)=f^{\prime}(s(t))(I-L)\mathbf{u}(t).
\]
Assuming that the eigenvectors of $L$ form a basis for $\mathbb{R}^{n}$, the
perturbations can be taken along an eigenvector of $L,$ $\mathbf{u}%
(t)=p_{i}(t)\mathbf{u}_{i}$, where $i\ge 2$ since the perturbations along
the direction $\mathbf{1}$ still yield a synchronous solution. The
amplitude $p_{i}(t)$ along the $i$th eigenvector obeys
\[
p_{i}(t+1)=f^{\prime}(s(t))(1-\lambda_{i})p_{i}(t) 
= p_{i}(0)\prod_{k=0}^{t}f^{\prime}(s(k))(1-\lambda_{i}).
\]
Thus, the system synchronizes if 
\begin{equation}
\lim_{t\rightarrow\infty}\prod_{k=0}^{t-1}\left|  f^{\prime}(s(k))||1-\lambda
_{i}\right|  = 0, \qquad  i=2,\dots,n. \label{sync-lambda}%
\end{equation}
In view of (\ref{stab-a}) and (\ref{lyap}), a sufficient
condition for local synchronization is
\begin{equation}
\max\{|1-\lambda_{i}|:i=2,\dots,n\}<e^{-\mu}. \label{sync-cond2}%
\end{equation}

The significance of the simple condition (\ref{sync-cond2}) is twofold. 
Firstly, it separates
the effects of the local (isolated) dynamics given by $\mu$ from the effects of the
network structure given by the left-hand side. Therofore, 
an appropriate synchronizability measure for the network is
\begin{equation}
\sigma\triangleq\max\{|1-\lambda_{i}|:i=2,\dots,n\},\label{sync-measure}%
\end{equation}
smaller values of $\sigma$ yielding synchronization
for a larger class of functions $f$. 
Secondly, the role of the
network structure on synchronizability is characterized by the spectrum of the
Laplacian. The only assumption
about $L$ used above is the existence of $n$ linearly independent
eigenvectors, which is generically satisfied by
matrices in $\mathbb{R}^{n\times n}$. Hence, $\sigma$ can be used for
comparing general networks with respect to their synchronizability, including
directed and weighted ones, and even when they have
different sizes.

For undirected weighted networks the synchronizability
measure (\ref{sync-measure}) simplifies to%
\[
\sigma=\max\{|1-\lambda_{2}|,|1-\lambda_{\max}|\}
\]
where $\lambda_{\max}=\lambda_{n}$. Then three types of networks can be distinguished.

(a) All eigenvalues are less than or equal to $1$. In this case
synchronizability is determined solely by $\lambda_{2}$, a larger value
implying better synchronizability through the condition $\lambda_{2}%
>1-e^{-\mu}$.

(b) All eigenvalues are larger than $1$. In this case synchronizability is
determined solely by $\lambda_{\max}$, a smaller value implying better
synchronizability through the condition $\lambda_{\max}<1+e^{-\mu}$.

(c) $\lambda_{2}\leq1\leq\lambda_{\max}$. In this case synchronizability
depends on both $\lambda_{2}$ and $\lambda_{\max}$, higher values of
$\lambda_{2}$ and smaller values of $\lambda_{\max}$ implying better
synchronizability through the condition
\[
\frac{\lambda_{2}}{\lambda_{\max}}>\frac{1-e^{-\mu}}{1+e^{-\mu}}%
\]

Note that in a weighted network the eigenvalues contain information about the
connection strengths. So, cases (a) and (b) can also be viewed as weakly and
strongly coupled networks, respectively, whereas (c) can be thought of as the
case of intermediate coupling strength. In this setting, the eigenratio
$\lambda_{2}/\lambda_{\max}$ has been used as a numerical measure of the
synchronizability of networks. In all cases, the critical quantity here is
often $\lambda_{2}$ since it can be arbitrarily small, whereas $\lambda_{\max
}$ can be bounded in terms of the largest vertex degree, as seen from
(\ref{est-degree}).

\section{Structural limitations to synchronization}

\label{sec:isoperimetric} The synchronizability of the network is
directly related to the its spectral properties by (\ref{sync-cond2}). Can the
same be said about the statistical properties? We shall show that the answer
is negative in general, although in certain cases some useful information
about synchronizability can be obtained. Throughout this section we deal with
undirected networks.

Using (\ref{est-degree}), it is seen that
\begin{equation}
\frac{\lambda_{2}}{\lambda_{\max}}\leq\frac{d_{\min}}{d_{\max}}%
.\label{deg-ratio}%
\end{equation}
Therefore, a network whose smallest and largest degrees are very different is
a bad synchronizer. For example, scale-free networks have poorer
synchronizability in comparison to some other architectures, as observed in
\cite{Pecora05}. Note, however, that (\ref{deg-ratio}) does \emph{not} imply that 
a more homogeneous degree distribution always means better synchronizability.
(In fact, we later give examples where a higher degree homogeneity
results in worse synchronizability.)  
For one thing, (\ref{deg-ratio}) is
only an upper bound for the eigenratio. Moreover, the bound depends on the
extreme degrees, whereas degree homogeneity (defined as the standard deviation
of the degree distribution) is an average quantity, which may only loosely
depend on the extreme degrees in large networks. In the following, we will use
more sophisticated bounds on $\lambda_{2}$ and derive general structural
limitations on synchronization. We will show that the effect of small
structural changes on synchronization need \emph{not }average out within the
large network structure, and therefore may not be captured by the average network properties.

For notation, let $V$ denote the set of vertices of an undirected network $G$,
and let $S\subset V$ be a subset of vertices, with $V-S$ denoting its
complement, and $|S|$ its cardinality. Define%
\[
|\partial S|=%
{\displaystyle\sum_{i\in S}}
\;%
{\displaystyle\sum_{j\in V-S}}
a_{ij}.
\]
In words, $|\partial S|$ is the (weighted) number of edges between $S$ and its
complement. The isoperimetric number\ $i(G)$ of a graph $G$ is defined by%
\begin{equation}
i(G)=\min\left\{  \frac{|\partial S|}{|S|}:S\subset V,\;0<|S|\leq\frac{n}%
{2}\right\}  .\label{h}%
\end{equation}
The computation of $i(G)$ is an NP-hard problem \cite{Mohar89}. However, an
important result in graph theory gives a lower bound for the isoperimetric
number in terms of the second eigenvalue of the Laplacian, namely,
$
i(G)\geq\frac{1}{2}\lambda_{2}.\label{h2}%
$
We turn the table around, and use this result and (\ref{h}) to
estimate  $\lambda_{2}\ $as%
\begin{equation}
\lambda_{2}\leq2\dfrac{|\partial S|}{|S|}\label{lambda-estimate}%
\end{equation}
where $S$ is any subset of vertices satisfying
$
0<|S|\leq n/2.\label{size}%
$

\begin{center}%
\begin{figure}
[tb]
\begin{center}
\includegraphics[
height=2.4007in,
width=2.9326in
]%
{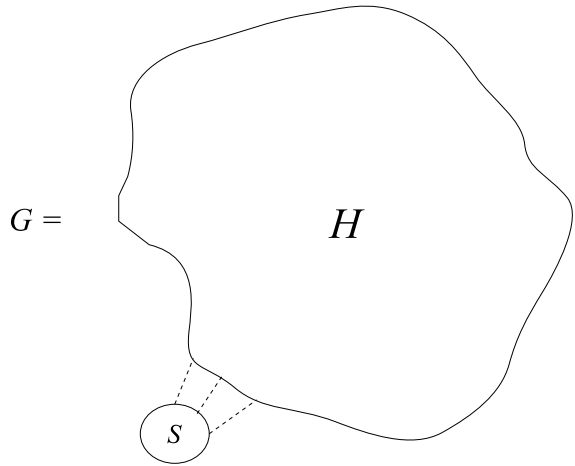}%
\caption{The statistical properties of the graph $G$ is determined by the huge
part $H$, while the value of $\lambda_{2}$ is independently constrained by the
small subgraph $S$.}%
\label{fig:graph-big}%
\end{center}
\end{figure}
\end{center}

The estimate (\ref{lambda-estimate}) holds the key to understanding why the
statistical properties of the network can fail to determine $\lambda_{2}$. The
important observation is that the bound on $\lambda_{2}$ is determined by the
properties of some subgraph $S$ and not in general by the graph itself. In
particular, $S$ can be very small compared to the whole
graph, in which case the statistical properties of the graph need not be
reflected in $S$, although the latter plays a crucial role in constraining the
value of $\lambda_{2}$. Figure \ref{fig:graph-big} illustrates the idea in
intuitive terms. Suppose in the graph $G$ we identify a huge part $H$ and a
much smaller part $S$. (Alternatively, we can imagine the possibility of
appending a small set of nodes $S$ to an existing graph, which is a
realistic scenario if one considers time-varying
connections which might come on and off 
\cite{Belykh04,Stilwell06,time-varying-sync}). By (\ref{lambda-estimate}), the
value of $\lambda_{2}$ is constrained by the properties of $S$. However, all
the gross statistical properties of $G$ are determined by $H$. If $H$ is any
graph which is claimed to have good synchronizability, we can force $G$ to
have poor synchronizability by appending $S$ to $H$. In other words, for large
networks, the synchronizability of $G$ and $H$ can be very different, although
many of their statistical properties are essentially the same. For instance,
if $S$ consists of 20 nodes and is connected to $H$ by one link, then by
(\ref{lambda-estimate}) $\lambda_{2}\leq0.1$ regardless of how $H$ is chosen.
Furthermore,  $\lambda_{2}/\lambda_{\max}$ can be considerably
smaller, especially if the average degree is high (viz. (\ref{est-degree2})),
which shows that a large average degree can actually impede synchronizability.
This example also illustrates the phenomenon observed in \cite{PRE05}; namely,
when two networks are combined by adding some links between them, the
synchronizability of the overall network decreases as the synchronizability of
individual networks is increased.

We have established that it is the local structures, described by the sets
$S$, that constrain the synchronizability, regardless of the global properties
of the network. Such local structures are conspicuous in certain types of
networks while they may not be so obvious in others. For example, the
situation shown in Figure~\ref{fig:graph-big} is typical for traffic or
transportation networks, where traffic is much denser within cities than
between them, and for interacting brain areas, where intracortical
connectivity is higher than inter-areal connections. We note, however, that
such local structural constraints need not exist in every network. For
instance, if the minimum degree is much larger than $n/2$, then the
ratio
$| \partial S|/|S|$ will be large for any subset $S$ satisfying $|S|\le n/2$.
However, such networks are very densely connected (the total
number of links being at least $n^{2}/4,$ which is about one half of 
that of a complete graph), whereas most real-world networks
are much sparser. Hence, if one considers networks which are not too densely
connected, it turns out that within essentially any family of graphs having
the same degree distribution, there exist graphs containing subsets $S$ for
which
$\vert$%
$\partial S|/|S|$ is small, and so the graph has a small second eigenvalue
$\lambda_{2}$. For a detailed mathematical proof the reader is referred
to~\cite{IEEE06}.

\begin{figure}
[ptb]
\begin{center}
\includegraphics[
height=1.7314in,
width=1.9424in
]%
{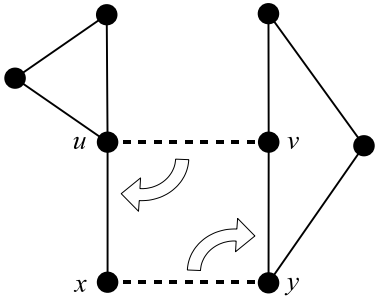}%
\caption{Link switching: The links $uv$ and $xy$ are deleted, and new links
$ux$ and $vy$ are added. The vertex degrees remain unchanged after the
operation. }%
\label{fig:linkswitch}%
\end{center}
\end{figure}

Without going into technical details, we here illustrate the essential ideas
to show that networks with the same degree distribution can have very
different synchronizability. A useful notion for this purpose is the use of
link switching to vary network properties without altering the vertex degrees
\cite{Edmonds64}. As depicted in Figure \ref{fig:linkswitch}, link switching
refers to the operation where, given two pairs of neighboring nodes $u,v$ and
$x,y$, one breaks the links $uv$ and $xy$ and replaces them by the links $ux$
and $vy$. The operation leaves the vertex degrees unchanged. As in the
particular case of Figure~\ref{fig:linkswitch}, the resulting network can be
disconnected, i.e. $\lambda_{2}$ becomes zero after the switch. Another
possibility is to link $u$ to $y$ and $v$ to $x$, which keeps the network
connected. It follows that $\lambda_{2}$ can be changed by link switching,
which makes it clear that the degree distribution does not determine
$\lambda_{2}$.%

\begin{figure}
[ptb]
\begin{center}
\includegraphics[
height=1.8075in,
width=2.2355in
]%
{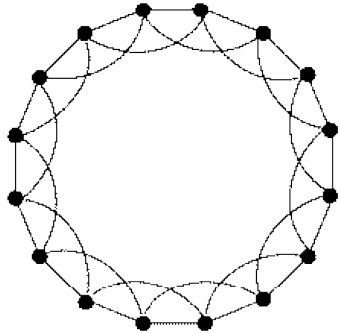}%
\caption{Circular arrangement of nodes.}%
\label{fig:circular}%
\end{center}
\end{figure}

\begin{figure}
[ptb]
\begin{center}
\includegraphics[
height=2.8253in,
width=4.03in
]%
{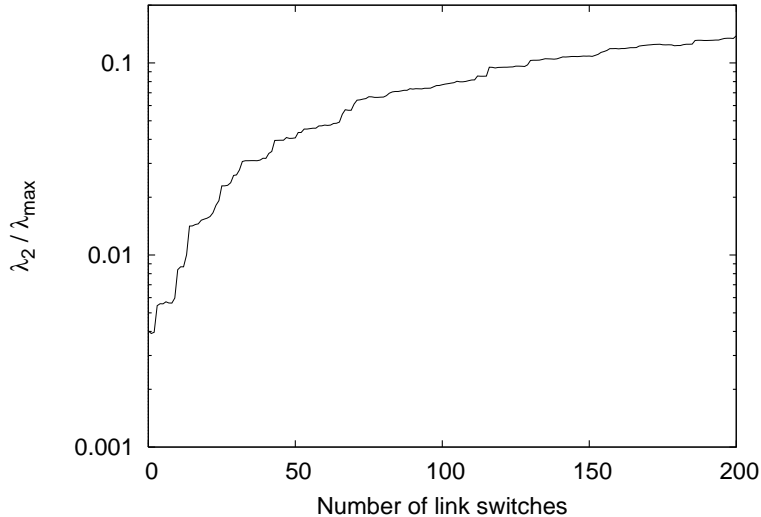}%
\caption{Improvement of synchronizability by random link switches, starting
from a circularly arranged regular network of 200 nodes with vertex degree
equal to 10.}%
\label{fig:switch}%
\end{center}
\end{figure}

As an example, we consider a circularly arranged set of nodes
where each node is connected to its $k$
nearest neighbors on each side (Figure~\ref{fig:circular}). 
Such circular structures have been heavily
used in numerical studies of the small-world effect, instigated by 
 \cite{Watts-Strogatz98}. We start with 200 nodes, 
where each node is connected to its 5 nearest neighbors
on each side, and randomly switch pairs of links so that the degree of each
node remains the same\footnote{The construction is similar to that in
\cite{Barahona02}, which adds random links to a circular arrangement of nodes,
whereas here we use link switching to keep the vertex degrees unchanged. }. It
is seen from Figure~\ref{fig:switch} that after only a few switches the
eigenratio increases by more than a factor of 10. In other words, the
circularly arranged network has very different synchronizability
characteristics than a typical regular network\footnote{A regular network 
is one where each vertex has the same degree.}, and leads to an
underestimation of the synchronizability for the latter. In fact, randomly
constructed large regular networks are typically expanders (see
Section~\ref{sec:discuss}), i.e., their eigenvalues $\lambda_{2}$ can be
bounded from below by a positive number.%

\begin{figure}[ptb]
\begin{center}%
\begin{tabular}
[c]{cc}%
{\parbox[b]{1.1147in}{\begin{center}
\includegraphics[
height=1.3111in,
width=1.1147in
]{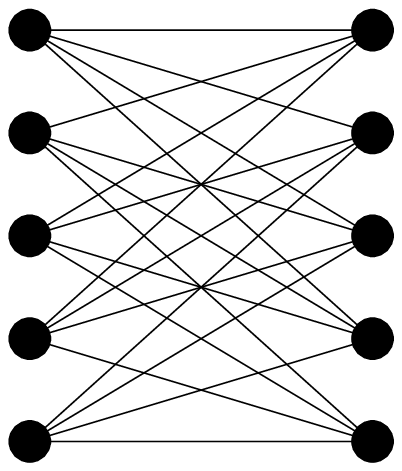}\end{center}}} & {\parbox[b]{1.4857in}{\begin{center}
\includegraphics[
height=0.7048in,
width=1.4857in
]{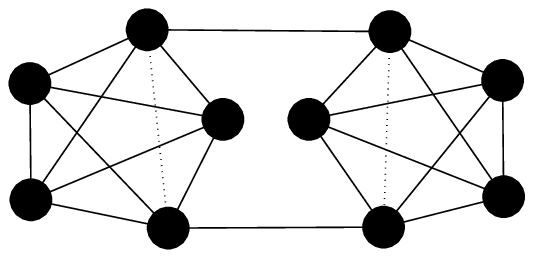}\end{center}}}\\
$G_{1}$ & $G_{2}$%
\end{tabular}
\end{center}
\caption{Two graphs with the same degree distribution and maximal degree
homogeneity, but very different synchronizability.}%
\label{fig:twographs}%
\end{figure}

\begin{figure}
[ptb]
\begin{center}
\includegraphics[
height=2.4751in,
width=3.5293in
]%
{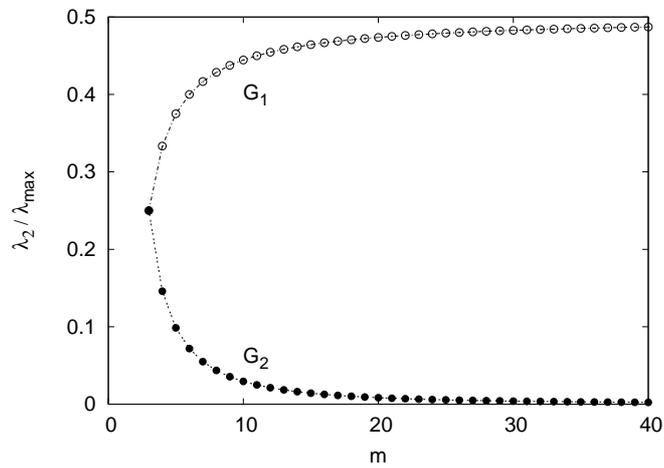}%
\caption{The eigenratio for the two graphs of Figure \ref{fig:twographs}%
.}%
\label{fig:twoeigenratio}%
\end{center}
\end{figure}

We next give a concrete construction for obtaining very good and very bad
synchronizing networks having identical vertex degrees. Consider regular
networks of $n=2m$ nodes where each node has degree $m-1$ \footnote{The degree
$k$ of each node can also be smaller than $m-1$ without changing the
subsequent argument. However, for $k>m-1$ the construction for $G_{2}$ fails.
Nevertheless, an average degree of $m-1$ for $2m$ nodes already implies a
well-connected network, the number of links $m(m-1)$ being about one half of
that of a complete graph, $m(2m-1).$ Since real networks are usually much
sparser than that, it suffices to consider $k\leq m-1$.}. We separate the
nodes into two groups of $m$ elements, and distribute the $m(m-1)$ links in
two different ways, as shown in Figure~\ref{fig:twographs}. In the first graph
$G_{1}$, all the links are across the two groups, and there are no connections
within a group. In mathematical terms, its adjacency matrix is given by%
\[
A_{1}=\left[
\begin{array}
[c]{cc}%
0 & J_{m}-I_{m}\\
J_{m}-I_{m} & 0
\end{array}
\right]  ,
\]
where $J_{m}$ denotes the $m\times m$ matrix whose every element is 1, and
$I_{m}$ is the $m\times m$ identity matrix. In the second graph $G_{2}$, we
start by putting all the links within a group, ending up with a disconnected
network with two components (the two pentagonal shapes in
Figure~\ref{fig:twographs}). The corresponding adjacency matrix is
\[
\quad A_{2}=\left[
\begin{array}
[c]{cc}%
J_{m}-I_{m} & 0\\
0 & J_{m}-I_{m}%
\end{array}
\right]  .
\]
To obtain a connected network, we use link switching to replace one link
within each component (shown by dotted lines) with a link connecting the two
components. In this way, for each $m$ we construct two different graphs with
the same degree distribution, and having the maximum homogeneity of vertex
degrees, since each one is a regular network. However, these two networks have
completely different synchronizability characteristics. Indeed, the first
network $G_{1}$ is related to the so-called complete bipartite graph. (If each
vertex degree were $m$ we would have exactly a complete
bipartite graph, in which case $\lambda_{2}=m=\lambda_{\max}/2.$) The
eigenvalues for $G_{1}$ are $\lambda_{2}=m-2$ and $\lambda_{\max}=2(m-1);$ so
the ratio $\lambda_{2}/\lambda_{\max}$ increases and tends to 1/2 as $m$ gets
large. For the second network $G_{2}$, we use (\ref{lambda-estimate}) to
estimate $\lambda_{2}/\lambda_{\max}\leq\lambda_{2}\leq4/m,$ which tends to
zero as $m$ gets large. Figure \ref{fig:twoeigenratio} shows the ratio
$\lambda_{2}/\lambda_{\max}$ for the two networks. It can be seen that a whole
range $(0,0.5)$ of values for $\lambda_{2}/\lambda_{\max}$ can be generated
using only regular graphs, which include very good as well as very bad
synchronizers. Furthermore, since all these graphs have maximally homogeneous
degree distribution, it is clear that the homogeneity of the degree
distribution does not determine synchronizability.

The argument above also shows that the average degree fails to determine
synchronizability: The average degree $m-1$ of both networks increases
with $m$; however, this increase results in a better synchronizability for
$G_{1}$ and a worse synchronizability for $G_{2}$. For a similar example which
uses non-regular networks, see \cite{PRE05}.

In closing this section, we mention that there are alternative definitions of 
the Laplacian, given by $I-D^{-1/2}AD^{-1/2}$ or $I-D^{-1}A$. The foregoing 
arguments apply also to the second eigenvalue of these matrices, with the same 
conclusions; see \cite{IEEE06}. Also for directed networks, one can use the 
idea of Fig.~\ref{fig:graph-big} to show that appending a small set of vertices $S$ 
to an existing graph disrupts synchronizability without affecting average 
statistical properties of the graph too much. Indeed, it is always possible to 
choose an $S$ having as few as two vertices and containing no directed spanning 
tree. Since the resulting graph $G$ also contains no directed spanning tree,
it is incapable of chaotic synchronization \cite{Wu-Nonlin05}.

\section{Discussion and conclusion}

\label{sec:discuss} The second eigenvalue $\lambda_{2}$ of the Laplacian is an
important invariant for undirected graphs. Also called the 
\emph{algebraic connectivity} or the \emph{spectral gap}, 
it has a special place within the Laplacian spectrum, and is deeply
related to many structural graph properties. For example, it comes up in
random walks on graphs, and consequently in epidemic spreading, as well as
robustness against edge and vertex removal (cut problems). Hence, the result
that $\lambda_{2}$ is not controlled by statistical properties of the network
has significance that goes beyond synchronization.

It is known from graph theory that there do not exist generally useful lower
bounds for the eigenvalue $\lambda_{2}.$ Some estimates can be obtained
asymptotically and in a probabilistic sense, i.e., almost surely as the
network size goes to infinity, and have been applied 
to study the asymptotic
behavior of synchronizability in power-law networks \cite{Kocarev05}. 
A nice mathematical result derived in \cite{Chung-PNAS03} for 
undirected random networks with given expected degrees
states that
\begin{equation}
\max_{i\geq2}|1-\lambda_{i}|\leq(1+o(1))\frac{4}{\sqrt{w_{\mathrm{avg}}}%
}+\frac{g(n)\log^{2}n}{w_{\min}}\label{chung}%
\end{equation}
where $w_{\mathrm{avg}}$ and $w_{\min}$ are the expected values of the average
and minimum degrees, respectively, $n$ is the network size, and $g(n)$ is some
slow-growing function of $n$. Note that the left-hand side of (\ref{chung}) is
the precisely the network synchronizability measure $\sigma$ defined in
(\ref{sync-measure}). The implication is that, 
under conditions that the second term on the right is
negligible compared to the first, one has essentially%
\begin{equation}
\sigma\leq\frac{4}{\sqrt{w_{\mathrm{avg}}}}\label{asymptotic}%
\qquad \text{and} \qquad        
\frac{\lambda_{2}}{\lambda_{\max}}\geq\frac{1-4/\sqrt{w_{\mathrm{avg}}}%
}{1+4/\sqrt{w_{\mathrm{avg}}}}.
\end{equation}
as $n\rightarrow\infty.$ 
These estimates in turn would imply that in the asymptotic limit 
a typical network with a large
average degree (namely, $\sqrt{w_{\mathrm{avg}}}$ at least as large as
$4e^{\mu}$ by (\ref{sync-cond2})) is a good synchronizer.
However, some care is needed in using (\ref{chung}) to derive conclusions about 
finite graphs. Since the $o(1)$ term has no bounds in terms of
graph size, it need not be small for a large but finite graph.
Moreover, to neglect the second term on the right-hand side of (\ref{chung}),
it is necessary that the expected minimum degree $w_{\min}$ grow faster than
$\log^{2}n$ as $n\rightarrow\infty$. In other words, (\ref{asymptotic}) can be
justified only for graphs for which both the size and the minimum degree are
very large. Unfortunately, real networks of interest are both finite and
sparse, and as our results indicate, (\ref{asymptotic}) does not necessarily
hold for these networks. In fact, (\ref{asymptotic}) is false even when we
restrict ourselves to a smaller class of graphs by imposing additional
restrictions in terms of network statistics, such as fixing the degree
distribution or requiring high degree homogeneity. Any such class would still
contain a bad synchronizer with positive probability. This is the precise
meaning of our statement that statistical properties do not suffice to
determine synchronizability.

Clearly, the true criteria for classifying networks with respect to
synchronizability involve the Laplacian spectrum, using some measure such as
(\ref{sync-cond2})\footnote{It should however be kept in mind that
(\ref{sync-cond2}) is a sufficient but not a necessary condition for
synchronization of chaotic systems.}. A closely related notion in graph theory
is that of expander graphs. Informally, these are families of sparse graphs
with high connectivity, so that the isoperimetric number defined by (\ref{h})
is bounded from below by some positive number. In this context, the
isoperimetric number $i(G)$ is also called the \emph{expansion constant}.
Recall that our arguments for identifying poorly-synchronizing networks are
based on showing that $i(G)$ can be small. Hence, in terms of the expansion
properties of networks, our results imply that the gross statistical
properties of networks do not suffice to characterize expander families.

There are many factors that contribute to the difficulties of studying complex
networks. The number of different networks of size $n$ increases dramatically
with $n$, which already makes it hard to obtain the relations between the
numerous network properties based on numerical simulations alone. One might
contend that numerical simulations give information about ``typical'' networks
in a certain class, but the mathematical proof of such assertions remains an
open problem. Moreover, the probability distributions from which
networks with specific statistical properties are drawn have rarely been
specified in the literature. On the other hand, a more subtle question worth
consideration is whether ``typical'' networks or properties carry all the
information that one should be interested in. The question is more meaningful
in the context of the complex networks found in nature, such as the human
brain or metabolic networks, which have very distinct functions and have
evolved after a long period of time into their present state. It is a
certainly intriguing possibility that their function may be related to, say,
their degree distribution. However, it is hardly warranted to claim that the
function is \emph{only} a consequence of that particular distribution. Such a
claim would imply that all networks with the same degree distribution are
similar at the functional level, which downplays the role of the millions of
years of evolution behind natural networks. In fact, one could argue that many
natural networks may be necessarily ``atypical'' in certain sense, if
evolution points along the direction of some optimization process. Hence, when
considering networks with such unique functions, the relevant features may be
those that make the network distinguished rather than typical, within the
considered class of networks.

In conclusion, dynamical processes on networks, and in particular
synchronization, are intimately related to the eigenvalues of the coupling
operator. As shown in the present paper, the gross statistical properties of
networks do not generally suffice to determine the spectrum. In other words,
the eigenvalues are among the intrinsic network features which determine the
dynamics and which are not derivable from the statistical characteristics.
Consequently, the spectral network properties deserve more attention in the
basic description and study of complex networks.

\end{document}